\documentclass[pra,groupedaddress,showpacs,dvips]{revtex4}

\usepackage{graphicx}

\newcommand\op[1]{\hat{#1}}     
\newcommand\hc[1]{{#1}^\dagger} 
\newcommand\ket[1]{|{#1}\rangle} 
\newcommand\bra[1]{\langle{#1}|} 
\newcommand\project[2]{ \ket{#1}\bra{#2} } 
\newcommand\ev[1]{\left\langle{#1}\right\rangle} 

\newcommand\conj[1]{{#1}^\ast}
\newcommand\abs[1]{|#1|}

\newcommand\eqn[1]{Equation~\ref{eq:#1}}
\newcommand\fig[1]{Figure~\ref{fig:#1}}

\newcommand\D\delta             
\newcommand\E{E}                
\newcommand\zero{\text{\sf 0}} 
\newcommand\one{\text{\sf 1}} 
\renewcommand\a{p}              
\renewcommand\b{q}              
\newcommand\z{z}                
\newcommand\A{\op{A}}                
\newcommand\B{\op{B}}
\newcommand\Z{\op{Z}}
\renewcommand\c{\op{c}}                
\renewcommand\H{\op{H}}
\newcommand\omegares\omega
\newcommand\delomega{\delta\omega}
\newcommand\deriv[1]{\frac{d#1}{dt}}

\newcommand\select[1]{\project{#1}{#1}}
\newcommand\half{{\textstyle\frac12}}
\newcommand\nbar{\bar{n}}

\newcommand\noise[1]{%
  (\nbar + 1) {\cal D}[c] {#1}
  + \nbar  {\cal D}[\hc{c}] {#1}}

\begin{document}

\title{Single electron measurements with a micromechanical resonator}

\author{R. E. S.~Polkinghorne}
\email{polkinghorne@physics.uq.edu.au}
\thanks{Work supported by the SRC for Quantum Computing Technology}
\author{G.~J.~Milburn}
\email{milburn@physics.uq.edu.au}
\homepage{http://www.physics.uq.edu.au/people/milburn/}
\affiliation{Centre for Quantum Computer Technology \\
  University of Queensland St. Lucia, Queensland, Australia}

\date{\today}

\begin{abstract}
  A mechanical electroscope based on a change in the resonant
  frequency of a cantilever one micron in size in the presence of
  charge has recently been fabricated.  We derive the decoherence rate
  of a charge superpositon during measurement with such a device using
  a master equation theory adapted from quantum optics.  We also
  investigate the information produced by such a measurement, using a
  quantum trajectory approach.  Such instruments could be used in
  mesoscopic electronic systems, and future solid state quantum
  computers, so it is useful to know how they behave when used to
  measure quantum superpositions of charge.
\end{abstract}

\maketitle

\section{Introduction}

As devices for processing and storing information become smaller the
demands on the readout technology become ever greater.  This is
especially true for proposed solid state quantum computers which store
information in various quantum degrees of freedom (qubits): in quantum
dots\cite{ld98}, nuclear spin\cite{k98,pvk98}, superconducting
islands\cite{npt99} and persistent currents\cite{fpctl2000}, to
cite just a small sample.  

Kane has proposed storing a qubit in the spin of a single phosphorous
nucleus implanted in silicon.  In his original readout scheme, this
was coupled by the hyperfine interaction to the spin of the donor
electron bound weakly to the nucleus.  A surface gate would then draw
the electron towards an adjacent ancilla donor, to which it might
tunnel, producing a doubly charged $D^-$ state.  Under appropriate
bias conditions, this transfer can only occur if the nuclear spin of
the qubit is oriented opposite the ancilla.  

A spin measurement is thus reduced to detecting the transfer of a
single electron charge to the ancilla.  This can be done by a
sensitive electroscope such as a single electron
transistor\cite{kmdcmsw2000}.  However, the techniques used for
fabricating microelectronics have recently been adapted to build
mechanical structures at micron and even nanometre
scales\cite{r2000}, and mechanical electroscopes sensitive to small
numbers of electrons have been constructed\cite{cr98}.  We will
consider how effectively such devices might perform the measurements
required for quantum information processing.

Classical treatments of measurement sensitivity assume that the
observable being measured has a definite value, which influences the
measuring instrument in a definite way.  The only question is how much
data we must gather to reliably distinguish this effect from other
influences on the apparatus, which produce noise.  Once we know the
size of the effect we wish to distinguish, and the level of noise in
the system, some elementary statistics tell us the integration time
required for a reliable measurement.

This assumption does not hold when we measure an observable of a
quantum system.  If the system is in a superposition state, the
observable will not have a definite value until some sort of
measurement is carried out.  Any interesting quantum information
device will produce such superpositions.  The process by which the
superposition is reduced so that the observable has a certain value
imposes a minimum level of noise in the measurement, which might be
increased by the same sources of technical noise that affect
measurements of classical systems.

In the proposed readout scheme for the Kane computer, a donor electron
is induced to tunnel between two phosphorous nuclei, depending on the
state of the nuclear spins.  In general, the nuclei are in a
superposition of a state which would permit tunneling, and one which
would prevent it.  After this tunneling has occured, the electon is
left in a superpostion of two position states, each localised on one
nucleus.  It then interacts with the electroscope, and in general with
other degrees of freedom in the crystal lattice, with the result that
we see it become localised on one nucleus or the other, so that the
electroscope gives a definite signal that the charge is present or
absent.

Note that we are not discussing an ensemble of quantum systems subject
to a single measurement, but rather a single quantum system subject to
a dynamical measurement process. In such a situation we need to be
able to describe the instantaneous conditional state of the measured
system as the measurement results accumulate.  This is quite different
from the usual situation that prevails in condensed matter systems,
where typically a measurement is made on a large number of (almost)
identical constituents undergoing quantum dynamics, and the
measurement results are already an average over an ensemble.
Fortunately mathematical techniques (known as quantum trajectory
methods) are available to describe the conditional dynamics of a
single quantum system subject to measurement with added noise, and
these methods have been applied with considerable success to
experiments in quantum optics and ion traps\cite{pk98}.  Recently such
methods have been applied to mesoscopic electronic
systems\cite{wusmkdc2001,gmws2001,k99a}.

\section{The Mechanical Electroscope}
\label{sec:electroscope}

\begin{figure}[h]
  \begin{center}
    \includegraphics{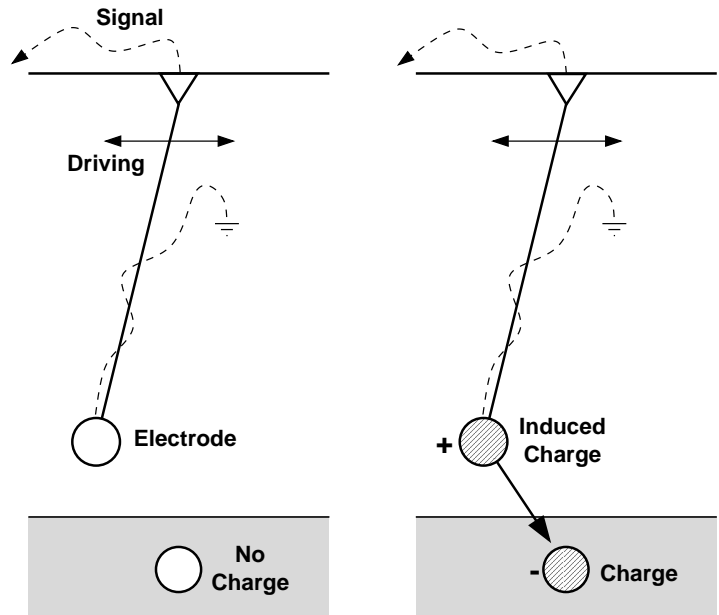}
    \caption{%
      Operation of the mechanical electroscope.  A charge trapped near
      the surface of some material is coupled to a cantilever
      suspended above the surface, as explained in the text.  The
      cantilever is driven at a rate $\E$ and damped by a combination
      of mechanical friction and reaction from the electronic readout
      loop at a rate $\gamma$.  If an excess charge is present on the
      surface, the frequency of the pendulum is increased by $\D$.
      For simplicity, the figure shows a simple pendulum, but in
      practice the cantilever would be a torsional pendulum,
      oscillating due to strain in the material.}
    \label{fig:diagram}
  \end{center}
\end{figure}

The operation of a micro-mechanical electroscope is shown
schematically in \fig{diagram}.  The active part is an electrode,
mounted on a cantilever no longer than $1\,\mu\text{m}$, which is set
in motion near the charge to be measured.  The electrode is held at
constant potential, so that its motion with respect to the unknown
charge induces a flow of charge between it and its voltage source.
The induced charge gives the electroscope electric potential energy as
well as elastic, and changes its resonant frequency.  If we envision
the electroscope being used to readout a qubit in a quantum computer,
there will be two charge states we wish to distinguish.  We will
denote the difference between the resonant frequencies of the
cantilever in these two states by $\delomega$; it is determined by
geometry and the mutual capacitance between the electrode and the
measured charge distribution.

We will assume the mechanical motion of the cantilever is elastic and
treat it as a simple harmonic oscillator.  Then its motion, including
the capacitive coupling to the target charge, is described by a
harmonic oscillator Hamiltonian
\begin{equation}
  H = \hbar ( \omega_0 + \delomega \, n_1 ) \hc{c}c
\end{equation}
where $\omega_0$ is the resonant frequency of the cantilever in the
absence of surface charge, and $c$ the annihilation operator for its
oscillation.  The observable $n_1$ will be defined shortly.


During readout of a Kane computer, a single donor electron may occupy
a bound state around either of two adjacent nuclei.  We will denote
these distinct spatial states by $\ket\psi$ and $\ket\phi$.  Only one
state (suppose $\ket\psi$) couples to the electroscope - this is how
we can distinguish them.

During readout, the surface gates will be configured to produce
tunneling between the two nuclei, depending on the state of the
nuclear spin qubits.  This entangles the charge states with the qubit
states $\ket\uparrow$ and $\ket\downarrow$.  We will denote the
combined states by \( \ket\zero = \ket\uparrow\otimes\ket\phi \), and
\( \ket\one = \ket\downarrow\otimes\ket\psi \), according to the
number of electrons interacting with the electroscope, which we will
represent by the operator \( n_1 = \project\one\one \).  In general the
measured qubit will be in a superposition state, so the total state
will take the form
\begin{equation}
\ket\Psi = a\ket\zero + b\ket\one
\end{equation}

Table~\ref{tab:data} gives numerical parameters for a cantilever
electroscope fabricated in 1998.  The frequency and operating
temperature of this electroscope meant that themal noise completely
dominated any quantum effects.  Besides lowering the temperature, this
could be changed by using a cantilever with a higher resonant
frequency, and such devices have been fabricated.  However, the
sensitivity of the electroscope depends on the frequency changing
significantly when change is present, and this might not be the case
in higher frequency cantilevers.


We note that the interaction Hamiltonian commutes with the number
operator $\hat{n}_1$.  Furthermore, in the absence of tunnelling, the
free Hamiltonian for the charge state itself is proportional to the
square of the charge (capacitive electrostatic energy) and itself
commutes with the charge number operator.  In the presence of the
measurement the number operator is thus a constant of motion. Such a
measurement is known as a quantum nondemolition
measurement\cite{wam}.  Number eigenstates are not changed by
the coupling to the apparatus, and moments of the number operator are
constant in time.  On the other hand any state that is initially a
coherent superposition in the number basis will be reduced to a
mixture diagonal in this basis, a process known as decoherence.  In an
ideal quantum nondemolition measurement, the probability distribution
for observed results at the conclusion of the measurement should
accurately reflect the intrinsic probability distributions of the
quantum nondemolition variable in the quantum state at the start of
the measurement.

This model, where the electroscope performs a QND measurement of the
coupled charge, is idealised.  If such an electroscope was used to
measure any interesting device, the motion of the cantilever would
disrupt the distribution of charge being measured.  The nature and
extent of this disruption would depend on the electrical properties of
the system being measured; for the Kane computer, determining these is
an unsolved problem in atomic physics.  In general, back action (and
interference from sources unrelated to the measurement) imposes a time
limit on the measurement, after which the charge state will have been
disrupted and the results will be meaningless.  The results of this
paper determine whether the electroscope can measure the charge with
the necessary precision within that time.


To detect the change in resonant frequency, we must set the cantilever
in motion with some driving mechanism.  In the device described in
Table~\ref{tab:data}, this was supplied by driving an alternating
current through a wire on the cantilever in the presence of a magnetic
field.  The current induced by the field in another wire was used to
monitor the response of the cantilever to the driving.

However, the details of the driving are not important.  As long as the
cantilever is coupled weakly to the driving system and is not damped
so strongly that its state changes significantly over the period of
its vibration (in other words, it has high finesse), the effect can be
described by a Hamiltonian.  In the interaction picture this takes
the form \( \hbar\mathcal{E}(\c+\hc\c) \), where $\mathcal{E}$ is the
strength of the driving in units of frequency.  If the finess of the
cantilever is low, noise from the driving system affects its motion
significantly, and the dynamics due to the driving can not be
approximated by a Hamiltonian.

The frequency shift could be detected in a number of ways. We could
sweep the driving frequency and monitor the amplitude of the
oscillations.  Or else we could drive the oscillator at a constant
frequency $\omega$, and then detect the change in phase of the
oscillation due to the shift in resonance frequency when a small
charge is coupled; this is the method analysed in this paper.  We will
assume that if the charge state is $\ket\zero$, the cantilever will be
driven on resonance; if it is $\ket\one$, the change $\delta\omega$ in
its resonant frequency will cause its phase to differ from that of the
driving force.  The rate of change of the phase of the output current
with frequency of the driving is greatest when the cantilever is
driven near its resonant frequency.

We will measure time by the inverse damping rate $\gamma^{-1}$.  Then,
defining a dimensionless driving strength \( E = \mathcal{E}/\gamma \)
and a detuning \( \Delta = \delomega/\gamma \), the Hamiltonian for the
coherent driving, in the interaction picture, is
\begin{equation}
  \label{eq:hamiltonian}
  \H_D = \hbar E \left( \c + \hc\c \right)
       + \hbar \Delta n_1\,\hc\c\c
\end{equation}

In reality of course the mechanical oscillations of the cantilever
will be subject to frictional damping, and accompanying mechanical
noise. The rate of energy dissipation is specified by the quality
factor, $Q$, which is the ratio of the resonance frequency to the
width of the resonance. For linear response, this gives
$Q=\omega_0/\gamma_M$, where $\gamma_M$ is the decay rate of energy
due to mechanical dissipation.  Roukes et al.\cite{cr99} have measured
quality factors up to $2 \times 10^4$. With such quality factors and
resonance frequencies approaching GHz, these devices are approaching
low quality optical resonators.  So we will treat the effect of
mechanical damping with the master equation methods of quantum optics.
These methods assume that the coupling of the resonator to the
dissipative degrees of freedom is sufficiently weak\cite{hr85,hrsw86}.
Specifically we assume that \( \gamma_M \ll \omega_0,\ kT/\hbar \).

Under these assumptions the coupling between the oscillator and the thermal
mechanical reservoir is\cite{gardiner}
\begin{equation}
  H_M= \sqrt{\gamma_M}(c\, a^\dagger(t)+c^\dagger\, a(t))
  \label{eq:thermal-noise}
\end{equation}
where $a(t),\ a^\dagger(t)$ are bosonic reservoir operators.  The
state of the reservoir will be taken to be that of a Planck thermal
equilibrium density operator with temperature $T_M$.

We now consider in more detail the mechanism by which the small
changes in resonance frequency induced by the proximity of a target
charge are transduced.  This may be done\cite{cr98} by fabricating a
wire loop on the mechanical oscillator and placing the whole apparatus
in a strong magnetic field. As the mechanical oscillator moves, an
induced EMF is set up in the loop and we may measure the induced
current.  When the current for the driving circuit is such as to drive
the mechanical oscillator at its resonance frequency, the induction
current is out of phase with the driving current. However when a small
target charge shifts the resonance frequency of the oscillator, the
induced current shifts in phase with respect to the driving current.
We can detect this phase shift by an electrical comparison of the
driving current and induction current. This is essentially homodyne
detection in which the driving current plays the role of a local
oscillator.  Unfortunately this electrical transduction of the
mechanical motion introduces another source of noise for the
measurement.

The induction current is coupled into an external amplifier circuit
which can be treated as a bosonic reservoir, with some non zero noise
temperature\cite{louisell}, $T_E$.  The readout circuit variable
coupled to the cantilever is the current operator $i(t)$ in the
readout circuit.  We will assume that the coupling is linear in the
current and coordinate degree of freedom of the cantilever.  
Under standard assumptions the interaction between the mechanical
oscillator and the readout circuit is described by the interaction
picture Hamiltonian,
\begin{equation}
  H_R = i \sqrt{\gamma_E}\,
    \left( c^\dagger \Gamma(t) - c \Gamma^\dagger(t) \right)
  \label{eq:elec-noise} 
\end{equation}
where $\Gamma(t)=b(t)e^{i\omega_0 t}$ with the actual current in the
circuit given by 
\( i(t) = \sqrt{\hbar\omega_0 / 2Lz_0} 
          ( b(t) + \hc{b}(t) ) \) 
, $L$ being the inductance per unit length of the transmission line, and
$z_0$ the quantisation length.  We will assume that the readout
circuit reservoir is bosonic and also in thermal equilibrium at some
temperature $T_E$.

Using the interaction Hamiltonians for the reservoir coupling
(Equations~\ref{eq:thermal-noise} and~\ref{eq:elec-noise}), we may obtain
the Heisenberg equations of motion for the oscillator and reservoir
variables. Using standard techniques\cite{gardiner}, the reservoir
variables may be eliminated to give a quantum Langevin stochastic
differential equation describing the dynamics of the oscillator
amplitude
\begin{equation}
  \frac{da}{dt} =
    -i \delomega \, a 
    -i\mathcal{E} - \frac{\gamma_M}{2}\,a - \frac{\gamma_E}{2}\,a
    + \sqrt{\gamma_M}\,a_{\text{in}}(t)
    +\sqrt{\gamma_E}\,b_{\text{in}}(t)
  \label{eq:qsde}
\end{equation}
%
where $a_{\text{in}}(t)$, $b_{\text{in}}(t)$ are the quantum noise
sources for the mechanical and electrical reservoirs respectively.
These noise terms are defined by correlation functions, which are
Fourier transforms of
\begin{eqnarray}
\ev{a_{\text{in}}(t)} & = & \langle b_{\text{in}}(t)\rangle = 0 \\
\ev{a^\dagger_{\text{in}}(\omega) a_{\text{in}}(\omega)^\prime} 
  & = & \bar{n}(\omega,T_M) \, \delta(\omega - \omega^\prime) \\
\ev{a_{\text{in}}(\omega) a^\dagger_{\text{in}}(\omega)^\prime}
  & = & ( \bar{n}(\omega,T_M) + 1 ) \, \delta(\omega - \omega^\prime) \\
\ev{b^\dagger_{\text{in}}(\omega) b_{\text{in}}(\omega)^\prime} 
  & = & \bar{n}(\omega,T_E) \, \delta(\omega - \omega^\prime) \\
\ev{b_{\text{in}}(\omega) b^\dagger_{\text{in}}(\omega)^\prime}
  & = & ( \bar{n}(\omega,T_E) + 1 ) \, \delta(\omega - \omega^\prime)
\end{eqnarray}
where
\begin{equation}
  \bar{n}(\omega,T) = \frac12 
    \left( \coth \left( \hbar\omega / 2 k_B T \right) - 1 \right)
\end{equation}
Note the equation explicitly includes a friction term (proportional to
$\gamma_E$) that arises form the electrical coupling to the readout
circuit.  The steady state average amplitude $\alpha_n=\langle
a(t)\rangle_{t\rightarrow \infty}$, is given by
\begin{equation}
  \label{eq:alpha}
  \alpha_n = \frac{-2i \mathcal{E}}%
                  {(\gamma_M+\gamma_E)+2i\,\delta\omega\,n_1}
\label{eq:ss-amp}
\end{equation}

\begin{figure}[h]
  \begin{center}
    \includegraphics{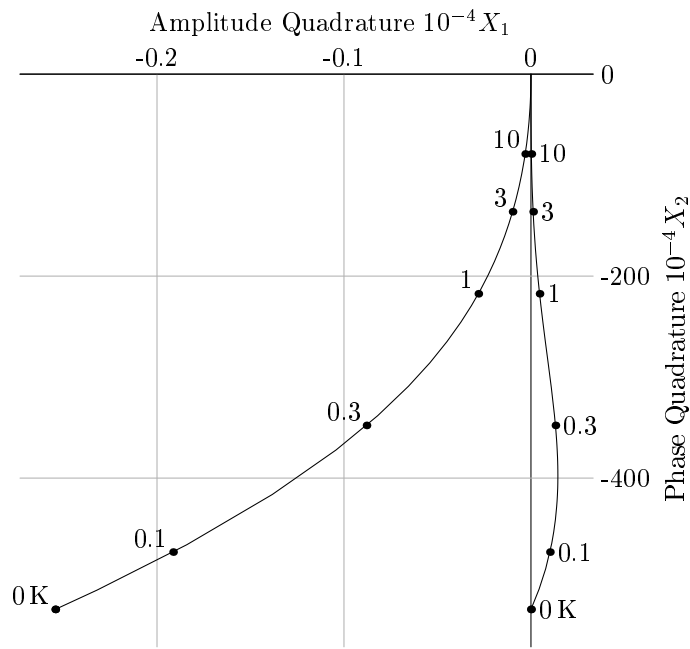}
    \caption{%
      The off-diagonal element $Z$ of the density operator is a
      thermal state displaced by amplitudes $\alpha$ and $\beta$,
      which depend on the temperature (see \eqn{Z}).  When the
      cantilever is coupled to a hot bath, these coherent amplitudes
      decrease, and $Z$ approaches a purely thermal state.  The values
      these amplitudes take in the Cleland and Roukes electroscope at
      temperatures from absolute zero up to $10\,\text{K}$ are plotted
      in the complex plane, in units of the ground state fluctuations.
      The amplitudes of the diagonal elements $A$ and $B$ do not vary
      with temperature, but remain at the $0\,\text{K}$ values.}
    \label{fig:amplitudes}
  \end{center}
\end{figure}

The actual measured quantity is the current in the readout circuit,
that is to say the readout variable is an electrical bath variable,
$b_{\text{out}}$ at the output from the system interaction. The output
amplitudes for both the mechanical and electrical baths are related to
the input variables for these two baths and the amplitude of the
mechanical oscillator by\cite{wam}
\begin{eqnarray}
  a_{\text{out}}(t) & = & \sqrt{\gamma_M}\,a(t) - a_{\text{in}}(t)\\
  b_{\text{out}}(t) & = & - i \sqrt{\gamma_E}\,a(t) - b_{\text{in}}(t)
  \label{eq:input-output}
\end{eqnarray}
The average value of the electrical readout amplitude in the steady state
is then found using equations \eqn{qsde} and \eqn{input-output}.
\begin{equation}
  \langle b_{\text{out}}\rangle = \sqrt{\gamma_E}\,\alpha_n
  \label{eq:output-amp}
\end{equation}
where $\alpha_n$ is given in \eqn{ss-amp}.  We see that the steady
state amplitude of the cantilever, and hence the output electrical
signal undergoes a change in phase and amplitude, see
\fig{amplitudes}. If we monitor the component in the imaginary
direction (that is, in quadrature with the driving signal, $E$) we
will have maximum sensitivity to this change in phase.  Furthermore it
is desirable to have $E$ as large as possible so that small changes in
phase translate into large changes in the quadrature.

We can now proceed to calculating the noise power spectrum for the
measured current.  The calculation is analogous to that for a
double-sided cavity given in reference\cite{wam}. We now do not
work in the rotating frame but return to the laboratory frame.  The
Fourier component of the output operator for the current is given by
\begin{equation}
  b_{\text{out}}(\omega) = 
    \frac{\left [\left (\frac{\gamma_E-\gamma_M}{2}\right )-i(\omega_0-
\omega)-i \,\delta\omega\, n_1 \right]b_{\text{in}}(\omega)
                                -i\sqrt{\gamma_E}{\cal
E}(\omega)+\sqrt{\gamma_E\gamma_M}a_{\text{in}}(\omega)}{\left [\left
(\frac{\gamma_E+\gamma_M}{2}\right )+i(\omega_0-\omega)+i \,\delta\omega\, n_1\right]}
\end{equation}
where ${\cal E}(\omega)$ is the Fourier component of the driving
amplitude. If the driving is noiseless and monochromatic, ${\cal
  E}(\omega)=\mathcal{E}\delta(\omega-\omega_d)$. However in reality there would
be some noise in the driving amplitude derived from the electrical
noise in the driving circuit. We will treat this as entirely
classical.

Equations \ref{eq:ss-amp} and \ref{eq:output-amp} suggest that the
signal will appear in the quadrature of the current out of phase with
the driving force, defined by
\begin{equation}
  \label{eq:quadrature}
  X_{2,out}(t) = i ( b^\dagger_{\text{out}}(t) - b_{\text{out}}(t) )
\end{equation}
with Fourier components $X_{2,out}(\omega)$. The measured power
spectrum is then given by the correlation function,
\begin{equation}
S_{2,out}(\omega,\omega^\prime)=\langle
X_{2,out}(\omega),X_{2,out}(\omega^\prime)\rangle
\end{equation}
Using the specified states for the electronic and mechanical noise
operators, we find,
\begin{equation}
S_{2,out}(\omega,\omega^\prime)=\left [|{\cal
B}(\omega)|^2(2 \bar{n}(\omega,T_E)+1)+|{\cal A}(\omega)|^2(2\bar{n}(\omega,T_M)+1)\right
]\delta(\omega-\omega^\prime)
\end{equation}
where
\begin{eqnarray*}
{\cal B}(\omega) & = & \frac{\frac{\gamma_E-\gamma_M}{2}-i\left ((\omega-
\omega_0)+ \,\delta\omega\, n_1\right )}
{\frac{\gamma_E+\gamma_M}{2}+i\left ((\omega-\omega_0)+
    \,\delta\omega\, n_1\right)}\\
 {\cal A}(\omega) & = &
\frac{\sqrt{\gamma_E\gamma_M}}{\frac{\gamma_E+\gamma_M}{2}+i\left((\omega-\omega_0)+
    \,\delta\omega\, n_1\right )}
\end{eqnarray*} 
To estimate the signal to noise ratio (SNR) we evaluate the spectrum
at the driving frequency (that is to say, at the central Fourier
component of the coherent driving);
\begin{equation}
S(\omega_0) = \frac{ \left[
      \left(\frac{\gamma_E-\gamma_M}{2} \right)^2 + (\delta\omega\, n_1)^2
    \right] (2 \bar{n}(\omega,T_E)+1)
    + \gamma_E\gamma_M(2 \bar{n}(\omega,T_M)+1)}
  {\left (\frac{\gamma_M+\gamma_E}{2}\right)^2 + (\delta\omega\, n_1)^2}
\end{equation}
Equations~\ref{eq:alpha}, \ref{eq:output-amp}
and~\ref{eq:quadrature} show that the magnitude of the Fourier
component of the mean signal at the driving frequency is given by
\begin{equation}
  |\langle X_{2,out}(\omega_D)\rangle| = 
    \frac{8 \sqrt{\gamma_E} \mathcal{E} \,\delta\omega\, n_1}%
         {(\gamma_M + \gamma_E)^2 + 4 \,\delta\omega^2\, n_1}
\end{equation}
The signal is a sharp peak at \( \omega = \omega_d = \omega_0 \), in
which there is a noise power \( S(\omega_0) \) per root Hertz.
So the SNR per root Hertz is \( |\langle X_{2,out}(\omega_D)\rangle|^2 /
 S(\omega_0) \), or
\begin{equation}
\text{SNR} = \frac
  {16 \,\gamma_E\, \mathcal{E}^2 \,\delta\omega^2\, n_1}
  { \left[ (\gamma_M + \gamma_E)^2 + 4 \,\delta\omega^2\, n_1 \right] 
    \left[
      ( \gamma_E-\gamma_M )^2 + 4 \,\delta\omega^2\, n_1
    \right] (2 \bar{n}(\omega,T_E)+1)
    + \gamma_E\gamma_M(2 \bar{n}(\omega,T_M)+1)}
\end{equation}
If the SNR required for the measurement is $\text{SNR}_r$, then we
must average over noise for a time $t$ such that \( \text{SNR}_r =
\text{SNR}/\sqrt{t} \).  If we set \( n_1 = 1 \), so we are measuring
the charge on one electron, the sensitivity is then \( e \sqrt{t} = e \,
\text{SNR}_r / \text{SNR} \).

\section{Unconditional description of the measurement.}

When we measure a quantum system, we bring an extremely large set of
independent observables of our instrument and its environment into
correlation with the measured system observable.  The environment of
the electroscope has two distinct components.  Firstly there is the
environment associated with the mechanical oscillator, which is
responsible for mechanical damping and noise. Secondly there is the
environment associated with the electrical readout, which is
responsible for Johnson-Nyquist noise in the electrical circuit, and
ultimately provides the measured result.  However, we are interested
in what the measurement tells us about the system, not in the exact
quantum state of the instrument and its environment.  Useful
instruments must operate independently of the detailed state of their
environments.

There are two ways to describe the partial state of the charge and
oscillator. Firstly we can ignore the results of the measurement and
average over states of the environment completely.  In this case the
evolution of the charge and oscillator is described by a master
equation.  Effectively we are averaging over the ensemble of partial
states distinguished by different measurement records

Secondly, we can ask for the conditional states of the charge and
oscillator, given a particular measurement record.  Each member of the
ensemble of partial states is associated with a distinct measurement
record of the instrument.  For it to be an effective measurement,
observers must be able to distinguish the states of the instrument.
In other words the charge must end up correlated with some simple
macroscopic quantity, like the current in a wire or the position of a
pointer on a scale.  It is then possible to ask for the particular
partial state of the measured system that is correlated with a known
pointer value. In other words we need to be able to specify the
conditional state of the system given a readout of the instrument
variable that distinguishes different charge states. This is the
conditional, or selective, description of the measured system.  Of
course if we average over the readout variables, we must obtain the
unconditional description of the system.

We begin with the unconditional description of the measurement.  The
dominant sources of excess noise that limit the quality of the
measurement are the thermal mechanical noise and thermal electrical
noise on the readout circuit.  Under certain Markoff and rotating wave
assumptions\cite{carmichael1,gardiner}, the explicit states of the
mechanical and electrical reservoirs may be traced out.  This leaves
the following master equation for the density operator of the
composite system of charge and cantilever,
\begin{eqnarray}
  \dot\rho(t) & = & 
    - i [ \mu (e n_1)^2 \, \hc\c\c , \rho] 
        - i [ \mathcal{E} \left( \c + \hc\c \right) , \rho] \\
    & & + \sum_{i = M,E} \gamma_i(\nbar_i + 1) {\cal D}[c] \rho
        + \gamma_i\nbar_i {\cal D}[c^\dagger]\rho
\end{eqnarray}
where  the superoperator $\cal D$ is defined by
\begin{equation}
  {\cal D}[c]\rho =
    c\rho\hc{c} - \frac{1}{2} ( \hc{c}c\rho + \rho\hc{c}c )
\end{equation}

This can be written in a more standard form
\begin{eqnarray}
  \label{eq:master}
  \dot\rho & = &
    - i [ \mu (e n_1)^2 \, \hc\c\c , \rho]
        - i [ \mathcal{E} \left( \c + \hc\c \right) , \rho] \\
    & & + \gamma (\nbar + 1) {\cal D}[c] \rho
        + \gamma \nbar  {\cal D}[\hc{c}] \rho
\end{eqnarray}
where \( \gamma \equiv \gamma_M + \gamma_E \), and
\( \nbar \equiv ( \gamma_M \bar{n}(\omega,T_M) + \gamma_E \bar{n}(\omega,T_E) )/\gamma \).

We will begin solving this master equation by separating the dynamics
of the cantilever and the charge.  As before, we assume there is only
one charge in the system, and consider the charge states $\ket\zero$
and $\ket\one$.  We can decompose $\rho$ into a $2 \times 2$ matrix of
cantilever operators
\begin{equation}
  \rho = \A \select\zero + \B \select\one
         + \Z\project\zero\one + \hc\Z\project\one\zero
  \label{eq:matrix}
\end{equation}
Since $\rho$ is Hermititan, we need only three cantilever operators,
$\A$, $\B$ and $\Z$.  We can now decompose \eqn{master} into three
independent equations involving only cantilever operators:
\begin{equation}
  \label{eq:no}
  \deriv\A = -i [ E(\c+\hc\c) , \A ] + \noise{\A}
\end{equation}
\begin{equation}
  \label{eq:yes}
  \deriv\B =
    -i [ E(\c+\hc\c) + \Delta \hc\c\c, \B ] + \noise{\B}
\end{equation}
\begin{equation}
  \label{eq:maybe}
  \deriv\Z =
    - i [ E(\c+\hc\c) , \Z ]
    + i \Delta \Z\hc\c\c + \noise{\Z}
\end{equation}
As before, we are now measuring time relative to the damping time $1/\gamma$.

If we measured the state of the charge by means other than the
cantilever, the state of the cantilever immediately after the
measurement would be $\B$ if the charge were present, or $\A$ if it
were absent.  Hence $\A$ and $\B$ must be density operators, and
Equations~\ref{eq:no} and~\ref{eq:yes} have the form of master
equations for a damped harmonic oscillator.  Such equations, and their
solutions, are familiar to quantum opticians.  The stable solution is
a displaced thermal state, which can be written
\begin{equation}
  \rho = \left( 1 - e^{-\lambda(t)} \right) \, D(\alpha(t)) \,
         e^{-\lambda(t)\hc\c\c} \, \hc{D}(\alpha(t))
\end{equation}
, where $D(\alpha)$ is a displacement operator \( \exp(\alpha\hc\c -
\conj\alpha\c) \), and in the steady state \( \lambda =
\hbar\omega_0/k_b T \).  In the limit of low temperature, \( kT \ll
\hbar\omega_0 \), this becomes a coherent state \( \select\alpha \).
In the steady state, the cantilever has as many thermal phonons as a
resevoir mode with the same frequency, i.e. \( e^{-\lambda} =
\nbar/(\nbar+1) \).  Its coherent amplitude $\alpha_0$ reaches a
balance with the driving and damping after a time around $2/\gamma$:
\begin{equation}
  \label{eq:amplitudes}
  \alpha(t) = \alpha_0 e^{-\kappa t/2}
    -\frac{2iE}\kappa (1 - e^{-\kappa t/2} )
\end{equation}
\begin{equation}
  \label{eq:cases}
  \kappa = \left\{
    \begin{array}{cc}
      1 & n = 0 \\
      1+2i\Delta & n = 1
    \end{array} \right.
\end{equation}
During measurement, the cantilever states $\A$ and $\B$ are
displaced thermal states with distinct coherent amplitudes.

As the measurement proceeds, we expect the charge state to evolve from
a coherent superpostion of $\ket\zero$ and $\ket\one$ to an incoherent
mixture; in terms of our decomposition, we expect the off-diagonal
term $Z$ to decay with time.  An operator of the form 
\begin{equation}
  \label{eq:Z}
  Z = \z(t) D(\alpha) \exp(-\lambda\hc\c\c) \hc{D}(\beta)
\end{equation}
where $\z(t)$ is a (possibly complex) amplitude, solves \eqn{maybe} if
$\alpha$, $\beta$, $\lambda$ and $z$ obey the following differential
equations:
\begin{equation}
  \label{eq:l}
  \deriv l = (\nbar+1) l^2 - (2\nbar + 1 + i\Delta) l + \nbar
\end{equation}
\begin{equation}
  \label{eq:a}
  \deriv a =
    \left( -i\Delta + (\nbar+1)l - \nbar - \half \right) a
    - iE(1-l)
\end{equation}
\begin{equation}
  \label{eq:b}
  \deriv b =
    \left( (\nbar+1)l - \nbar - \half \right) b
    + iE(1-l)
\end{equation}
\begin{equation}
  \label{eq:k}
  \deriv k = -iE(a - b) + (\nbar+1)(l + ab - 1) + 1
\end{equation}
Here \( l = \exp(-\lambda) \), \( a = \alpha - l\beta
\), \( b = \conj\beta - l\conj\alpha \), and \( k = \log
z + l\conj\alpha\beta - \half(\abs\alpha^2 + \abs\beta^2) \)

In general, these equations can be solved numerically.  However, there
are some special cases where we can get interesting information
analytically.  First we consider the zero temperature limit, where the
off diagonal term $\Z$ is a projector \( z \project\alpha\beta \).
The amplitudes $\alpha$ and $\beta$ are the amplitudes of the diagonal
terms given by \eqn{cases}, and $z$ is a complex amplitude.  Once
$\alpha$ and $\beta$ have reached their steady state, the trace of the
off-diagonal term, decays exponentially with a rate \(
\abs{\alpha-\beta}^2/2 \).

If we assume the detuning $\Delta$ is small, and hence \(
\abs\beta \approx \abs\alpha = 2E \), The difference between the
steady state amplitudes of \eqn{cases} is
\begin{equation}
  \abs{\alpha - \beta}^2 
    = \frac{16 E^2 \Delta^2}{1+4\Delta^2}
    \approx 4 \abs\alpha^2 \Delta^2
\end{equation}
Cleland and Roukes give enough information about their devices for us
to calculate this explicitly \cite{cr98}.  Using the data in
Table~\ref{tab:data}, we can calculate $\alpha$ from the definition of
the annihilation operator for a torsional pendulum
\begin{equation}
  \alpha = \ev\c 
         = \sqrt{\frac{\kappa}{2\hbar\omega_0}} \ev{\theta_{\text{max}}}
         = 5.3 \times 10^6
\end{equation}

The normalised detuning can be
calculated from the frequency shift per electron and the measured
quality factor:
\begin{equation}
  \Delta = \delta\omega /\gamma = \frac{2 \pi \, \delta\nu \, Q}{\omega_0}
         = 2.4 \times 10^{-4}
\end{equation}
The decoherence rate is then \( 3.2 \times 10^{6} \, \gamma \), or \(
8.1 \times 10^{9} \, \text{s}^{-1} \).

\begin{figure}[h]
  \begin{center}
    \includegraphics{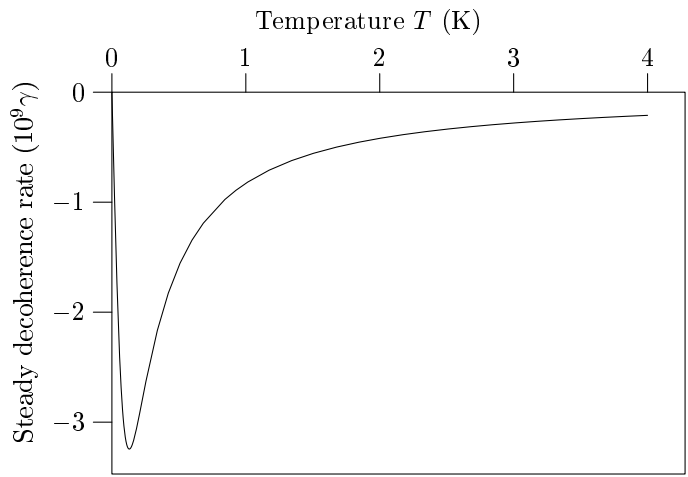}
    \caption{%
      When the measurement has been running for a time around
      $1/\gamma$, and the cantilever amplitudes have reached their
      steady state, any remaining coherence between the two charge
      states decays exponentially.  Here the rate of this decay is
      plotted as a function of temperature, for the device described
      in Table~\ref{tab:data}.  The maximum decay rate of $-3.2 \times
      10^9\,\gamma$ occurs at $130\,\text{mK}$.  Beyond this point the
      decay rate decreases with temperature, possibly because the
      increased thermal noise makes the coherent amplitude of the
      cantilever harder to distinguish.
      }
    \label{fig:graph}
  \end{center}
\end{figure}

\begin{figure}[h]
  \begin{center}
    \includegraphics{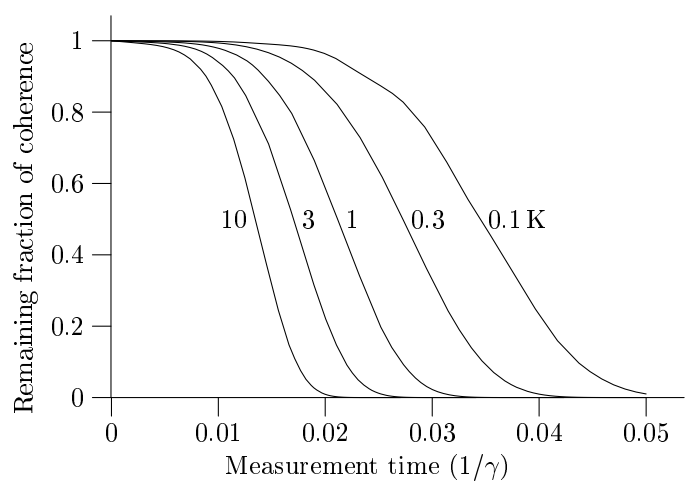}
    \caption{%
      The state of the cantilever takes some time to become entangled
      with the charge after they begin to interact, as the cantilever
      state moves towards its steady value.  After this the charge
      state decoheres rapidly.  Here the coherence between the two
      charge states is plotted as a function of time for an array of
      temperatures.  The cantilever is initially in a thermal state at
      the appropriate temperature.  Note the the charge state has
      decohered long before the cantilever reaches its steady
      amplitude, which occurs after a time $1/\gamma$.}
    \label{fig:coherencetime}
  \end{center}
\end{figure}

As $\nbar$ increases from zero, the amplitudes $\alpha$ and $\beta$
for the off diagonal operator $\Z$ are reduced, as shown in
\fig{amplitudes}.  The initial decay of $z(t)$ is shown in
\fig{coherencetime}, and the steady decay rate, i.e. the limit of \( |
z^\prime(t)/z(t) | \) when \( t \gg 1/\gamma \), in \fig{graph}.  At
low temperatures (below $130\,\text{mK}$), the increased thermal noise
from the bath causes $\Z$ to decay more rapidly as the temperature of
the bath is increased.  Contrary to expectations, the steady
decoherence rate of the charge superposition decreases as the bath
temperature increases above $130\,\text{mK}$.  The extra thermal noise
increases the overlap between the oscillator states corresponding to
the presence and absence of charge.

\section{Conditional description}
\label{sec:trajectories}

We now turn to the correlations between the charge and the resevoir
system.  These are important because we must be able to distinguish
the results corresponding to different charges to make a
measurement of the charge at all.  They can be studied most simply
using quantum trajectory theory, which associates charge states with
possible observed states of the apparatus \cite{carmichael-lect}.

We will assume we monitor the current in the electrical resevoir; this
is equivalent to an optical homodyne measurement \cite{sa96}.  The
inferred state of the charge as such a measurement proceeds is
governed by a Wiener process, which is generated by a stochastic
increment $dW$.  The average of $dW$ over the ensemble of possible
measurement results is zero.  Since the deviation of the Wiener
process represented by $dW$ increases proportional to $\sqrt t$, the
average of $(dW)^2$ is $dt$.  The simplest way to manipulate such
differentials is to modify the chain rule, to give what is know as Ito
calculus.

Given a particular measurement result, labelled by a Wiener increment
$dW$, the evolution of the charge and cantilever is
\begin{equation}
  d\ket\psi = \left( \frac1{i\hbar} \H dt
    - \frac\gamma2 
        \left( \hc{c}c - 2 \ev{\frac{x}2}c + \ev{\frac{x}2}^2 \right)dt
    + \sqrt\gamma \left( c - \ev{\frac{x}2} \right)dW \right) \ket\psi
\end{equation}
When we insert the charge and cantilever Hamiltonian, and normalise time
by the damping rate as before, this becomes
\begin{equation}
  \label{eq:trajectory}
  d\ket\psi = \left( - i( E(c+\hc{c}) + \Delta n\hc{c}c )dt
    - \frac12\left( \hc{c}c - 2 \ev{\frac{x}2}c + \ev{\frac{x}2}^2 \right)dt
    + \left( c - \ev{\frac{x}2} \right)dW \right) \ket\psi
\end{equation}
When a particular function $dW$ is selected from the Wiener ensemble,
this can be solved to show the evolution of a pure state $\ket\psi$.
These states form an ensemble with density operator $\rho$.  Of course
$\rho$ can be decomposed into many ensembles, so the evolution
generated by \eqn{trajectory} is not unique.  The details are given in
Carmichael\cite{carmichael-lect}.

Mixed states of the cantilever and charge must be written in the form
of \eqn{matrix}.  However pure states can always be written as
\begin{equation}
  \ket\psi = \ket{A} \otimes \ket\zero + \ket{B} \otimes \ket\one
\end{equation}
as before we will assume the state of the cantilever is initially
coherent, so that
\begin{equation}
  \ket\psi = p \ket{\alpha 0} + \b \ket{\beta 1}
\end{equation}
The differential of a scaled coherent state $\b(t) \ket{\beta(t)}$ is
\begin{equation}
  d (\b \ket{\beta})
    = \left(d \b - \frac12 \b d\abs\beta^2\right) \ket{\beta}
    + \b \dot\beta \, dt \, \hc{c} \ket{\beta}
\end{equation}
comparison with \eqn{trajectory} gives Equations~\ref{eq:amplitudes}
for the evolution of $\alpha$ and $\beta$ as before.  Some Ito
calculus manipulations show that
\begin{equation}
  d \abs{\b}^2 =
    \abs{\a \b}^2 (\ev{x}_\alpha - \ev{x}_\beta) dW
\end{equation}
where \( \ev{x}_\alpha \) is the expectation value of the amplitude
quadrature $x$ in a coherent state $\ket\alpha$, which is just \( 2
\text{Re}\,\alpha \).  The normalisation of $\ket\psi$ requires that
\( d \abs{\a}^2 = - d\abs{\b}^2 \).

We need to compare the gain in knowledge shown by this
trajectory picture to the decay of coherence modelled by the master
equation.  The results of the measurement are the probabilities
$\abs{\a}^2$ and $\abs{\b}^2$; the pure state which the observer
will infer from these has a density operator
\begin{equation}
  \label{eq:inferredrho}
  \rho = \abs{\a}^2 \project00 + \abs{\b}^2 \project11
    + \abs{\a \b} ( \project10 + \project01 )  
\end{equation}
The off-diagonal
terms in this have magnitude $\abs{\a \b}$; we can average over $dW$
to see the behaviour of the density operator for the ensemble of
measurement results.

Some more routine Ito calculus gives the evolution of this:
\begin{equation}
  d \abs{\a \b} = - \abs{\a \b}
    \left( \frac18 \left( \ev{x}_\beta - \ev{x}_\alpha \right)^2 dt
      + \frac12 \left( \abs{\a}^2 - \abs{\b}^2 \right)
        \left( \ev{x}_\beta - \ev{x}_\alpha \right) dW \right)
\end{equation}
Since the average of $dW$ over different measurement results is zero,
on average
\begin{equation}
  \label{eq:sensitivity}
  d \abs{\a \b} = \
    - \frac18 ( \ev{x}_\beta - \ev{x}_\alpha )^2 \abs{\a \b} dt
\end{equation}

If the difference between the charges associated with states $\ket0$
and $\ket1$ is $e$, then in the state \( \a \ket{\alpha 0} + \b
\ket{\beta 1} \), the uncertainty in the charge is given by
\begin{equation}
  \left(\ev{(ne)^2} - \ev{ne}^2\right)^\half = e \abs{\a \b}
\end{equation}
From \eqn{sensitivity}, this decreases exponentially as the
measurement progresses, at a rate
\begin{equation}
  \frac18 ( \ev{x}_\beta - \ev{x}_\alpha )^2 
    = \frac{ 8 \Delta^2 \E^2 }{ ( 1 + 4\Delta^2 )^2 }
\end{equation}
This differs from the square root decay of classical uncertainty as
measurements are averaged over time, but exponential decay is what we
would expect for decay of coherence\cite{wam}.  For the device
described in \cite{cr98}, this is almost equal to the decoherence
rate.  In real devices, thermal noise will cause the trajectory states
to be mixed, however the evolution of such mixed states is much harder
to calculate.

\section{Discussion}
\label{sec:discussion}

To estimate the time required for our measurement, we have calculated
how long it takes for an initially pure superposition of charge states
to be reduced to a mixture, and how long (in some sense) it takes us
to find out which charge eigenstate we have been left with.  While
these questions are interesting in their own right (they composed the
deepest mystery of physics for the best part of a century), it could
be argued that they don't reflect the way measurements would be used
in a real computer.

The most that we could do with measurements on pure states is state
preparation.  In a coherent quantum computer this would be rather
pointless though, since if we know the initial state we could just
rotate it into the eigenstate we want.  We carry out measurements to
find out something we don't know: in other words we apply them to
mixed states, with a view to finding out which of the possibilities is
real.

Information theory provides tools to quantify this, such as
conditional entropy and mutual information.  Unfortunately calculating
any of these requires knowledge of the ensemble of trajectories
generated by each component of the mixture, and the overlaps between
them.  In general it is hard to find the probability distribution of
trajectories; we usually just calculate averages.  It might be worth
doing this numerically, however.

There is a more straightforward limitation to our analysis: in present
day devices the thermal effects that we have neglected in the
trajectory treatment utterly dominate the vacuum noise we have
considered.  Hence the measurement time will be limited by the need to
average classical fluctuations.  It is possible that future devices
operating at higher frequencies will reduce the level of thermal noise
so that quantum effects will be important.  This presents the
remarkable prospect of a solid cantilever with position and momentum
known to the limit allowed by the uncertainty principle.


%
%

\begin{table}[h]
  \begin{center}
    \begin{tabular}{lcc}
      Operating temperature $T$ & $4.2\,\text{K}$
       & $k_{\text{B}} T = 3.6 \times 10^{-4}\,\text{eV}$ \\
      Resonant frequency $\omega_0/2\pi$ & $2.6\,\text{MHz}$ 
       & $\hbar\omega_0 = 1.1 \times 10^{-8}\,\text{eV}$ \\
      Torsional spring constant $\kappa$ 
       & $1.1\times 10^{-10}\,\text{Nm}$ & \\
      Amplitude $\theta_{\text{max}}$ & $30\,\text{mrad}$ & \\
      Frequency shift per electron $\delta\nu$
       & $0.1\,\text{Hz}$ & \\
      Quality factor $\omega_0/\gamma$ & $6.5\times 10^3$ &
    \end{tabular}
    \caption{Data for an electroscope fabricated by Cleland and Roukes
      \cite{cr98}.}
    \label{tab:data}
  \end{center}
\end{table}

\end{document}